# Improving Students' Understanding of Lock-In Amplifiers

Seth DeVore, Alexandre Gauthier, Jeremy Levy, and Chandralekha Singh

*Department of Physics and Astronomy, University of Pittsburgh, Pittsburgh, PA 15260*

**Abstract:** A lock-in amplifier is a versatile instrument frequently used in physics research. However, many students struggle with the basic operating principles of a lock-in amplifier which can lead to a variety of difficulties. To improve students' understanding, we have been developing and evaluating a research-based tutorial which makes use of a computer simulation of a lock-in amplifier. The tutorial is based on a field-tested approach in which students realize their difficulties after predicting the outcome of simulated experiments involving a lock-in amplifier and check their predictions using the simulated lock-in amplifier. Then, the tutorial guides and helps students develop a coherent understanding of the basics of a lock-in amplifier. The tutorial development involved interviews with physics faculty members and graduate students and iteration of many versions of the tutorial with professors and graduate students. The student difficulties and the development and assessment of the research-based tutorial are discussed.

## INTRODUCTION

The lock-in amplifier (LIA) is an instrument used extensively in laboratory research, especially in condensed matter physics [1-3]. However, many students who use this instrument for their research have only a limited understanding of the operation of LIAs. Often, the LIA is used as a "black box" to find the amplitude of a signal at a given frequency without understanding the instrument's internal workings. Improper use of LIAs, and misinterpretation of data obtained from them, are unfortunately quite common.

We investigated the common difficulties that graduate students have with lock-in amplifiers in their experimental research. Based on these difficulties, we developed a research-based "OnRamp" tutorial to ease the transition of those who are just beginning their research in the lab setting, as well as to provide a firmer foundation for those who already use LIAs in their research. The OnRamp focuses on helping students build a robust understanding of the fundamental operation of a LIA, and helps students develop an intuitive feel for many of the possible situations that they may encounter in their experiments.

Many graduate students make use of the LIA in conjunction with an optical "chopper" or some type of modulator, e.g., an amplitude modulator in a laser. In this instance, many of the more complex behavior of a LIA are bypassed, leading the student to believe (incorrectly) that they are qualified to use a LIA in other contexts. But even with a "chopping" experiment, the choices of chopper frequency and location can greatly affect the result. Specifically, the choice of chopping frequency, and its proximity to noise sources (e.g., 50 Hz or 60 Hz "line" noise) can greatly affect the quality of the data obtained. Strategies for reducing noise also depend where the noise enters into the signal stream [3]. Having a robust understanding of the LIA is therefore critical to making decisions regarding experimental design.

Our goal in developing this research-based tutorial was to develop tools that can instill an intuitive understanding of the basics of the LIA functions, so that students who use LIAs in their research understand more deeply how the input parameters affect the output. By merging conceptual and mathematical aspects of the instrument, the tutorial strives to help students learn the relationship between the input-settings-and expected outputs so that they are able to troubleshoot unexpected outputs in their laboratory work.

## THE IDEAL LOCK-IN AMPLIFIER

Throughout this paper, as in the tutorial, we will treat the LIA as an idealized version of the instrument. The most common use of LIAs in the lab is to measure small signals in the presence of large background noise. Here we assume that the signal of interest is centered around a frequency $f_S$. In general it will not be a pure frequency since the amplitude can change, and amplitude modulation leads to sidebands that surround the central frequency. To separate the signal of interest from unwanted noise, a reference signal is defined. The reference signal is selected to have a unit amplitude (for convenience) and a frequency $f_R$ equal to the frequency which the experimentalist wants to analyze in the input signal (that is to say, $f_R = f_S$). The (idealized) single-frequency input signal is first pre-amplified by a factor $g$, to give $V_I = gA_S\cos(2\pi f_S t + \varphi)$. This amplified signal is then multiplied (or "mixed") by a reference signal $V_{RX} = \cos(2\pi f_R t)$ to form the "unfiltered" x-channel output of the LIA: $V_{MX}(t) = V_I(t)V_{RX}(t)$. Here, $\varphi$ is the phase of the input signal of frequency $f_S$ with respect to the reference signal, and $A_S$ is the amplitude of the input signal with frequency $f_S$. A similar process is followed to produce the unfiltered y-channel: $V_{MY}(t) = V_I(t)V_{RY}(t)$, where $V_{RY} = \sin(2\pi f_R t)$.

To understand the effect of this multiplication we rely on a trigonometric identity to reveal: $V_{MX} = V_I V_{RX} = \frac{1}{2}gA_S[\cos(2\pi(f_S - f_R)t + \varphi) + \cos(2\pi(f_S + f_R)t + \varphi)]$ and $V_{MY} = V_I V_{RY} = \frac{1}{2}gA_S[\sin(2\pi(f_S - f_R)t + \varphi) - \sin(2\pi(f_S + f_R)t + \varphi)]$ Generally speaking, $f_S - f_R \ll f_R$ and $f_S + f_R \approx 2f_R$. Finally, $V_{MX}$ and $V_{MY}$ are fed through two low-pass filters with a "time constant" $\tau = f_0^{-1}$ and "rolloff" (expressed in dB/octave), which should be chosen carefully based upon the nature of the experiment. Signals of frequency $f \ll f_0$ are passed with unity gain, while signals with $f \gg f_0$ are attenuated as $f^{-n}$ for $6n$ dB/octave filters (e.g., $\propto f^{-2}$ for 12 dB/octave filters). This filtering yields the two approximately time-independent output signals $V_{OutX}$ and $V_{OutY}$.

For the case where $f_S = f_R$ precisely, the low-pass filter will remove the second-harmonic ($2f_R$) term from both $V_{MX}$ and $V_{MY}$, providing information about the magnitude and phase of the input signal: $V_X = \frac{1}{2}gA_S\cos\varphi$, and $V_Y = \frac{1}{2}gA_S\sin\varphi$, or $A_S = 2/g\sqrt{V_X^2 + V_Y^2}$, and $\varphi = \tan^{-1}(V_Y/V_X)$.

## METHODOLOGY

We individually interviewed five physics professors at the University of Pittsburgh (Pitt) who conduct research in CMP and commonly work with graduate students who use LIAs in their research. A typical interview time was 60-90 minutes, during which each professor was also asked to articulate what he expected his students to know about LIAs, what was the goal of the professor's experiment(s) utilizing LIAs and how LIAs are useful in the broader framework of research. While two co-authors of this manuscript conduct research in experimental condensed-matter physics, the other two had an opportunity to use an actual LIA in a research laboratory. Using the feedback from professors as a guide along with a cognitive task analysis of the underlying knowledge involved in the operation of a LIA, we developed a preliminary tutorial along with a pre-test and post-test (to be given before and after the tutorial, respectively). We also developed a LIA simulation, which forms an integral part of the tutorial.

We then interviewed graduate students using a think-aloud protocol to better understand their difficulties and to fine-tune the tutorial. In these semi-structured interviews, students were asked to talk aloud while they worked through the pre-test, tutorial and the post-test. The interviewer tried not to disturb students' thought processes while they answered the questions except to encourage them to keep talking if they became quiet for a long time. Later, the interviewer asked students for clarification of points they had not made clear earlier in order to understand their thought processes better. Some of these questions were planned out ahead of time while others were emergent queries based upon a particular student's responses to questions during an interview.

The tutorial (along with the pre-test and post-test) was iteratively refined 18 times, based upon feedback from graduate students and professors. While professors worked through the different versions of the tutorial and associated simulations at their convenience and provided feedback afterward in one on one meetings, we used a think-aloud protocol when graduate students worked on any version of the tutorial. Based upon feedback from professors and graduate students, we refined the tutorial. After the tutorial was fine-tuned to our satisfaction, it was administered to six physics graduate students who had not been involved in the development phase of the tutorial, but who either concurrently used a LIA for their research or had been exposed to one in the past. They were administered the pre-/post-tests before and after the tutorial in order to assess its effectiveness. Three researchers jointly deliberated the rubric for scoring performance on pre-/post-tests (see Table 1).

The final version of the tutorial helps students learn the purpose and function of the mixer as well as the low-pass filter, which are essential components of the LIA. The tutorial also helps them connect the conceptual and mathematical aspects of the operation of the instrument. Most questions require students to predict output signals based on given input parameter sets for the instrument, providing hints and feedback to help them with their predictions as needed [4]. Students are given an opportunity to check their predictions with a simulation that replicates the function of this instrument. These predictions and corresponding simulations cover a wide array of possible LIA settings and input signal types. For each prediction and simulation, both a mathematical and a more conceptual intuitive explanation of the output signal based on the input parameters are provided, in case the students cannot reconcile their predicted output with the output shown on the screen in the simulation. Moreover, in some questions, students are given information about the output signal of the LIA for hypothetical cases and asked questions about the corresponding input parameters. These questions too are supplemented with hints and feedback to guide their reasoning of the correct answer in each case.

**Table 1.** Summary of the grading rubrics used for the pre-test and post-test questions discussed.

| Case 1 Pre-test and Post-test | | | | | |
|---|---|---|---|---|---|
| Non-zero frequency present? | Yes | Magnitudes of both DC components (x,y) are zero? | Yes | 5 points | +2.5 points for correct frequency +2.5 points for correct amplitude |
| | | | No | 2.5 points | |
| | No | 0 points | | | |
| Case 2 Pre-test and Post-test | | | | | |
| Non-zero frequency present? | Yes | Magnitudes of both DC components (x,y) are correct? | Yes | 5 points | +2.5 points for correct frequency +2.5 points for correct amplitude |
| | | | No | 2.5 points | |
| | No | Magnitudes of both DC components (x,y) are correct? | Yes | 2.5 points | |
| | | | No | 0 points | |

## RESULTS

The difficulties found in our investigation have roots in a lack of coherent understanding of the fundamentals of a LIA. For example, students often had a fuzzy understanding of what the mixers in a LIA do. As noted, the mixer multiplies the input signal (frequency $f_s$) by the reference signal (frequency $f_R$). This results in sum and difference frequencies, $|f_S \pm f_R|$, appearing in the output of the mixer [3]. The special case $f_S = f_R$, which is commonly encountered by students in the lab, results in an output consisting of a DC (time-independent) signal and a signal with frequency $2f_S$. Of the six students interviewed, none made use of the relevant equations for the multiplication of the input and reference signals in the mixer to gain insight into what output they should expect in the questions posed in the pre-test. One interviewed student said: "I never realized that I didn't actually calculate these things [he did not realize that there is a mathematical framework that can be used to make sense of the output voltage for a given input parameter set]".

Interviews suggest that another aspect of the LIA that is often overlooked by graduate students is the effect of the low pass filter on the output signal. For example, interviewed students had great difficulty with the fact that the frequencies that will make it into the output signal can be estimated by making use of the time constant, $\tau$. The time constant is inversely proportional to the cutoff frequency which is the frequency at which the low-pass filter will cause half of the power in the signal input to it to be lost. Above this frequency, an increase in the frequency leads to increased signal attenuation.

Next, we discuss the details of the difficulties in two situations (both with $\varphi = 0$ and $g = 2$). The specifics of these difficulties and their underlying causes became evident through the process of performing student interviews. We analyze student performance on questions from the pre-test and post-test for each of these two cases. We note that the pre-test and post-test questions were matched for content as closely as possible but were not identical. If there was a difference (other than different values of parameters) between the corresponding pre-/post-test questions, the post-test was perceived by the researchers to be more difficult.

Case 1 involves $f_S$ being close to but not equal to $f_R$. The difficulty with Case 1 ($f_S \neq f_R$) was probed by a pre-test question in which $f_R = 199$ Hz, $f_S = 200$ Hz, $A_S = 2.5$ V and $\tau = 0.1$ s (the corresponding post-test question had $f_R = 249$ Hz, $f_S = 250$ Hz, $A_S = 2.0$ V and $\tau = 0.025$ s). Students were asked to predict the output signal. For both these pre-/post-test questions, a 1 Hz signal is present in the output, the only difference being that for the pre-test, the amplitude of the output signal will be 2.5 V and in the post-test, the amplitude will be 2.0 V.

In the pre-test, two of the six students interviewed claimed that all frequencies present in the signal that do not equal $f_R$ will be attenuated. They did not take into account the value of the time constant $\tau$ and how it determines the frequencies present in the output signal. When explicitly asked about the time constant, a student said: "The time constant is something that I don't always think about quantitatively, unless I absolutely need to". Table 2 shows that the average score obtained on the pre-test for Case 1 (using the rubric in Table 1) is 37.5% while the post-test average score is 91.7%. Only three of the six students realized that a non-zero frequency is present in the output signal in the pre-test while all of them recognized that a non-zero frequency is present in the output signal in the post-test. Discussions suggest that the tutorial helps students understand the function of the low pass filter and realize that the component at the difference frequency $|f_S - f_R|$ in the output of the mixer is present in the output signal for the given situation.

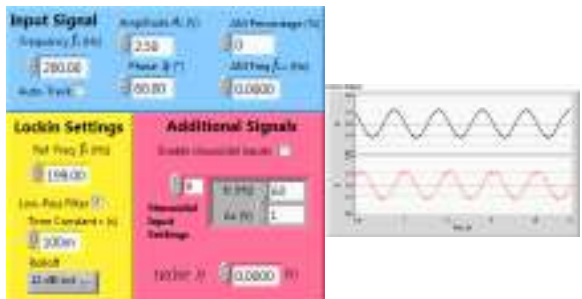

**FIGURE 1.** Screen capture of the input parameters (left) and the correct output signal (right) for Case 1 pretest question.

**TABLE 2.** Average score of six graduate students on the pre-/post-test for Case 1 and 2. Case 1 involves $f_S \neq f_R$ and Case 2 involves two frequencies in the input signal.

| Difficulty Present | Average Score on Pre-test Question | Average Score on Post-test Question |
|---|---|---|
| Case 1 | 37.5% | 91.7% |
| Case 2 | 0.0% | 75.0% |

Case 2 involves an input signal with two frequencies present. The difficulty with Case 2 was probed by the pre-test question in which $f_R = 1$ Hz, $f_S = 260$ Hz, $f_C$ (a second input signal frequency) $= 1$ Hz, $A_S = 1.0$ V, $A_C$ (the amplitude of the second input signal frequency) $= 3$ V, and $\tau = 0.025$ s. The post-test question had the following input parameters: $f_R = 250$ Hz, $f_S = 250$ Hz, $f_C = 253$ Hz, $A_S = 2.0$ V, $A_C = 1.0$ V, and $\tau = 0.02$ s. Students were then asked to predict the outputs. In the pre-test, the first component of the input signal is entirely attenuated and, therefore, has no effect on the output question leaving the output to be determined entirely by the two components due to the presence of the second input signal. This yields an output with a 1 V constant component and a sinusoidal component with a 2 Hz frequency and a 1 V amplitude. However, for the post-test question, both the first and second input signals enter into determining the output. For each input signal frequency, the difference frequency (with respect to the reference) is passed and the sum frequency is attenuated. This yields an output with a 2 V DC component and a sinusoidal component with a 3 Hz frequency and a 1 V amplitude. Incidentally, a second input signal frequency as in Case 2 can often be caused by a coherent noise, e.g., the 60 Hz signal from the power lines present in the lab. Interviews suggest that graduate students often assumed that any input signal fed into a LIA must have only one coherent frequency in addition to white noise, without considering factors such as a coherent noise. Without an in depth understanding of the fundamentals of the LIA, they had difficulty, especially in the pre-test, in predicting what kind of output one should expect when two frequencies are present in the input signal. Table 2 shows that the average scores for Case 2 are 0.0% and 75.0% on the pre-/post-test, respectively. This data shows that no student answered any part of the Case 2 pre-test question correctly. However four out of the six students answered the post-test question completely correctly. Interviews suggest that there is a distinct increase in their understanding of both multiple frequency signals and the effect of $\tau$ on the output signal since both effects must be understood to answer this question correctly.

## SUMMARY

We find that physics graduate students who use LIAs for their experimental research have many common difficulties with the basics of this instrument. We have developed and evaluated a research-based tutorial that helps students learn the basics of how this instrument operates and also helps students make connections with the pertinent mathematics that describes the operations of its major components.

## ACKNOWLEDGEMENTS

We thank National Science Foundation for award NSF-1124131.